%% file: main.tex
\documentclass[conference]{IEEEtran}
\usepackage{cite}
\usepackage{amsmath,amssymb,amsfonts}
\usepackage{algorithmic}
\usepackage{graphicx}
\usepackage{textcomp}
\usepackage{xcolor}
\usepackage{fancyhdr}
\usepackage[hyphens]{url}
\usepackage{comment}
\usepackage{xspace}
\usepackage[font=small]{caption}
\usepackage{enumitem}

\def\BibTeX{{\rm B\kern-.05em{\sc i\kern-.025em b}\kern-.08em
    T\kern-.1667em\lower.7ex\hbox{E}\kern-.125emX}}

\fancypagestyle{firstpage}{
  \fancyhf{}

  \fancyfoot[C]{\thepage}
}

\title{\vspace{-2ex}Practical Encrypted Computing for IoT Clients\vspace{-1ex}} 
\author{McKenzie van der Hagen\\Carnegie Mellon University\\Pittsburgh, U.S.A.\\mckenziv@andrew.cmu.edu\vspace{-3ex} \and Brandon Lucia\\Carnegie Mellon University\\Pittsburgh, U.S.A.\\blucia@andrew.cmu.edu\vspace{-3ex}}
\newcommand{\syslong}{Client-aided HE for Opaque Compute Offloading\xspace}
\newcommand{\sys}{CHOCO\xspace}
\newcommand{\arch}{CHOCO-TACO\xspace}
\newcommand{\archlong}{Client-aided HE for Opaque Compute Offloading Through Accelerated Cryptographic Operations\xspace}

\newcommand{\totalarea}{19.3 mm$^2$\xspace}

\newcommand{\encenergy}{.1228 mJ\xspace}

\newcommand{\encenergysave}{603$\times$\xspace}
\newcommand{\encmaxenergysave}{648$\times$\xspace}

\newcommand{\vggenergysave}{37\%\xspace}
\newcommand{\sqzenergyoverhead}{82.5\%\xspace}

\newcommand{\enctime}{.66 ms\xspace}

\newcommand{\encspeedup}{417$\times$\xspace}
\newcommand{\encmaxspeedup}{1094$\times$\xspace}
\newcommand{\decspeedup}{125$\times$\xspace}

\newcommand{\avgnnspeedup}{121$\times$\xspace}

\newcommand{\avgheaxtimeovhd}{25$\times$\xspace} 
\newcommand{\avgswspeedup}{$1.7\times$\xspace} 
\newcommand{\avghwvstflite}{$2.2\times$\xspace} 
\newcommand{\blavgoverhead}{24$\times$\xspace}
\newcommand{\tacovsheax}{$54.3\times$\xspace} 
\newcommand{\tacovssw}{$123.27\times$\xspace} 

\newcommand{\mpccommmin}{14$\times$\xspace} 
\newcommand{\mpccommmax}{2948$\times$\xspace} 

\begin{document}
\maketitle
\thispagestyle{firstpage}
\pagestyle{plain}


\begin{abstract}
Privacy and energy are primary concerns for sensor devices that offload
compute to a potentially untrusted edge server or cloud. Homomorphic
Encryption (HE) enables offload processing of encrypted data. 
HE offload processing retains data privacy, but is limited by the need for
frequent communication between the client device and the offload server.
Existing client-aided encrypted computing systems are optimized for
performance on the offload server, failing to sufficiently address client
costs, and precluding HE offload for low-resource (e.g., IoT) devices. 
We introduce \syslong (\sys), a client-optimized system for encrypted offload
processing. \sys introduces rotational redundancy, an algorithmic optimization
to minimize computing and communication costs.  We design \archlong (\arch), a
comprehensive architectural accelerator for client-side cryptographic
operations that eliminates most of their time and energy costs.
Our evaluation shows that \sys makes client-aided HE offloading feasible for
resource-constrained clients. Compared to existing encrypted computing
solutions, \sys reduces communication cost by up to \mpccommmax. With hardware
support, client-side encryption/decryption is faster by \encmaxspeedup and
uses \encmaxenergysave less energy. In our end-to-end implementation of a
large-scale DNN (VGG16), \sys uses \vggenergysave less energy than local
(unencrypted) computation.
\end{abstract}

\input{src/introduction}
\input{src/background}

\input{src/motivation}
\input{src/system}
\input{src/architecture}

\input{src/designspace}
\input{src/evaluation}
\input{src/relatedwork}
\input{src/conclusion}



\bibliography{refs}
\bibliographystyle{IEEEtran}

\end{document}

%% file: src/introduction.tex
\section{Introduction}\label{sec:introduction}

Data-producing client devices have a long history of decreasing in size and energy
storage capability~\cite{wisp,capy,amulet,flikker} leading the way to trillions of tiny devices~\cite{arm-iot-trillion}.  As clients scale down, the sophistication of computations on sensor data is scaling up, now often using complex machine learning (ML). With little total energy (potentially no battery) \cite{wisp, capy, amulet, flikker, permacam, camaroptera} and constrained compute and memory, a device is fundamentally limited in its local processing capability.

A contrast to local compute is ``inference as a service'' offloading.  A
server houses a large collection of DNN models and processes data from many 
clients, without consuming client memory and energy resources for processing.
Such centralized models are easy to evolve, requiring a single update to the
model on the server, and avoiding the need to re-distribute an updated model
to a large network of fielded client devices. 
{\em Data privacy} is the main barrier to realizing these benefits of offload
computing: offloading exposes sensitive user data to a shared, potentially
untrusted offload server. 

Recent work offers several options for privacy-preserving computation, including trusted execution environments (TEEs)\cite{SGX,
nestedSGX, SecureEnclaves}, differential privacy (DP), multi-party computation
(MPC)\cite{Soteria, XONN, EzPC}, and homomorphic encryption (HE) \cite{LoLa,
CHET, CryptoNets, nGraphHE2}.  
Client-aided, hybrid HE-MPC protocols have seen recent success for DNN inference\cite{Gazelle, Delphi, MiniONN, Cheetah}, owing to their ability to use HE to protect user data and process
linear operations (e.g. convolution) on encrypted data.
Hybrid HE-MPC protocols obfuscate intermediate results of linear operations
and send them to the client, which applies non-linear operations (e.g.
activations) using MPC.  HE-MPC imposes the high client-side cost of MPC to
ensure privacy, not only of client data, but also model data.  If an
application requires only client data privacy, then HE-MPC needlessly imposes
MPC's cost.


Hybrid HE-MPC implementations of privacy-preserving DNN inference show
promising results, but have largely neglected to address the added compute
burden on the client. Existing solutions optimize HE-MPC parameters to the
benefit of the centralized model server, for both performance and model
privacy. Systems choose large ciphertext sizes (MBs), requiring {\em
gigabytes} of client-server communication for a single inference. It is
infeasible for resource-constrained client devices, to participate in such
schemes. 


This work identifies the ``middle-ground'' between fully-local compute, with
its associated resource requirements and inability to use centrally-managed
models, and hybrid HE-MPC, with its prohibitive compute and communication
costs imposed on client devices for model privacy.  We propose \syslong
(\sys), a system for privacy-preserving computation that minimizes client
costs. \sys targets applications that do not require model privacy, but that
require strict client data privacy. \sys reduces client costs by {\em orders
of magnitude} over HE-MPC, availing resource-constrained client devices of the
benefits of privacy-preserving ML.

\sys is client-aided HE without MPC, performing encrypted linear
operations on the server and plaintext non-linear operations on the client.
\sys introduces {\em rotational redundancy}, a new encrypted permutation
algorithm that minimizes client communication and resource requirements.
Additionally, a unique facet of client-aided HE motivates \sys: in a typical
HE scheme, encryption and decryption happen once per computation, but in
client-aided HE, encryption and decryption happen repeatedly on the critical
path.  
We quantitatively show that the prohibitively high time and energy cost to
encrypt and decrypt is the client's primary bottleneck. We propose \archlong
(\arch), a {\em comprehensive} hardware accelerator implementing all of each
of these HE cryptographic primitives and virtually eliminating their time
cost.


Our evaluation of a complete hardware-software implementation demonstrates
the benefits of a client-optimized system for privacy-preserving computation.
Comparing to seven prior HE and/or MPC approaches, \sys reduces client communication
costs by {\em orders of magnitude}, with improvements ranging from
\mpccommmin--\mpccommmax.
\arch's hardware acceleration improves client time and energy by \tacovssw
compared to software and \tacovsheax compared to HEAX, which accelerates some
(but not all) cryptographic sub-operations in hardware.  Our results show
that \sys makes client-privacy-preserving DNN inference comparably performant
to local inference (using TFLite), sometimes even {\em exceeding} the
performance of local inference, while avoiding the limitations of
fully-local compute.  Our main contributions are:
\begin{itemize}
\item \sys, a client-optimized system for privacy-preserving computation enabling encrypted computing for resource-constrained devices.
\item Rotational redundancy, an encrypted permutation algorithm that minimizes client-server communication.
\item \arch, a specialized hardware accelerator for client-side HE primitives.
\item A full hardware-software implementation of client-aided
privacy-preserving DNN inference that improves client costs by orders of
magnitude compared to HE-MPC, performing comparably to local computation, while enjoying the benefits of model centralization. 
\end{itemize}

%% file: src/background.tex
\section{Background \& Motivation}\label{sec:background}
\sys allows an IoT device to offload computation to a more-capable server that
computes on encrypted data using Homomorphic Encryption (HE). Primitive operations, most of which support encrypted, Single Instruction
Multiple Data (SIMD) computation, are composed into HE algorithms.  These algorithms are in turn used to build
encrypted HE applications, such as DNN inference.

\subsection{Homomorphic Encryption}\label{sec:background:HE}
Homomorphic encryption is a class of cryptography schemes that allow computing
on encrypted data. 
In HE, for a pair of messages $m_1$ and $m_2$ that can be
manipulated by an operation $\oplus$, a homomorphic version of the
operation $\oplus'$, and encryption/decryption operations,
$Enc()$, $Dec()$,  the homomorphic operation applied to the encrypted data
produces a result that, when decrypted is equal to the operation applied to
unencrypted data:

\begin{equation}\label{eq:heproperty}
Dec(Enc(m_1) \oplus' Enc(m_2)) = m_1 \oplus m_2
\end{equation}

\begin{figure}[!ht]
\centering
\includegraphics[width=\linewidth]{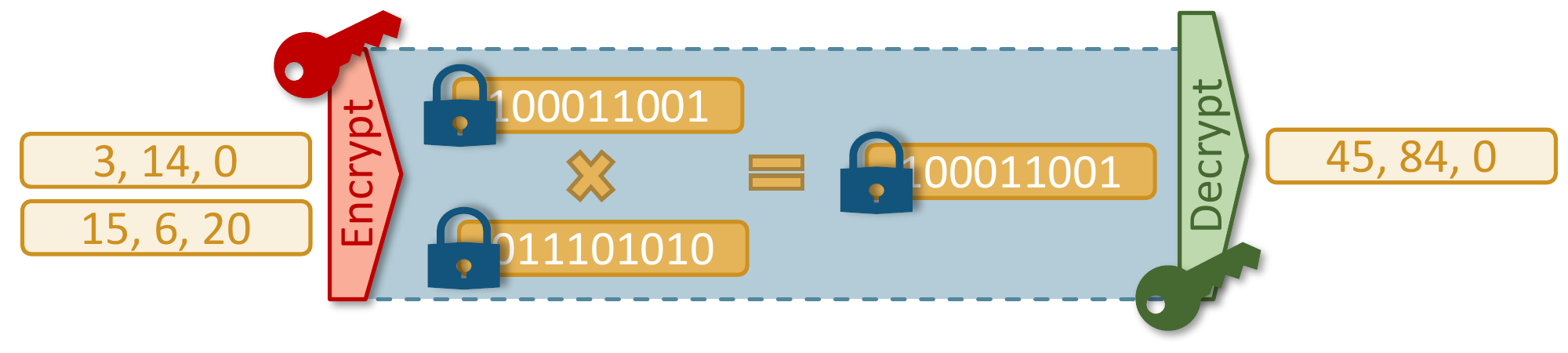}
\caption{Homomorphic Encryption allows for computation directly on encrypted vectors of data.}
\label{fig:fheexplain}
\end{figure}

Modern HE schemes~\cite{FV,ckks} are based on the ring learning with errors
problem (RLWE). These schemes encrypt a {\em vector} of thousands of values
into the coefficients of a large polynomial,  hiding the vector's contents
through modular arithmetic and the addition of noise. The HE operations in
Table~\ref{tab:ops} manipulate ciphertexts, producing new ciphertexts
containing the result of an operation applied element-wise to the input as
depicted in Figure \ref{fig:fheexplain}. Each operation adds a predictable
amount of noise to the encrypted vector, with some operations (e.g.,
multiplication) adding a large amount of noise, and others (e.g., addition)
adding little. 
The arithmetic depth of an HE computation is limited by noise growth.  A
sequence of operations that exhausts the noise budget renders decryption
impossible and data unrecoverable.  To avoid exceeding the noise budget, a
system must schedule encrypted operations to limit noise growth. 

\input{table/ops}

HE can {\em refresh} a ciphertext to eliminate accumulated noise, replenishing the full budget.
{\em Fully} Homomorphic Encryption (FHE) refreshes noise without decryption at
enormous computational cost through ``bootstrapping''~\cite{gentryFHE}. In
contrast, {\em Somewhat} Homomorphic Encryption
(SHE)~\cite{bfv1,bfv2,ckks,SEAL3.4,Gazelle} refreshes noise with pre-scheduled decryption and re-encryption operations.

\input{table/params}

\sys uses the increasingly practical Brakerski/Fan-Vercauteren (BFV) SHE scheme\cite{bfv1, bfv2} in SEAL~\cite{SEAL3.4}.
Table \ref{tab:params} summarizes the scheme's parameters, which 
dictate its security, computational complexity, noise budget,
and ciphertext size:

\begin{equation}\label{eq:ciphersize}
w \times N \times s \times (k-1)
\end{equation}

Typically, the polynomial modulus, $N$, is a power of two between $2^{11}$ and
$2^{15}$. A fresh ciphertext is two polynomials ($s = 2$) of $N$ elements each. For
a given $N$, a smaller coefficient modulus $q$ provides higher security but
a smaller noise budget. A practical $q$ value is hundreds of bits. Operating
directly on such large values is inefficient; HE schemes use the Residual
Number System (RNS)~\cite{rns} to represent numbers using $k$ smaller,
co-prime moduli. SEAL uses 60-bit residual moduli, to fit in a 64-bit machine
word.

BFV supports integer operations modulo the plaintext modulus $t$.  A larger
$t$ allows for larger numbers but contributes to a smaller noise budget. Parameter
selection is application-dependent and must allow for sufficiently large
plaintext values while maintaining a budget for sufficient noise growth. Table
\ref{tab:paramselect} shows \sys's parameters and ciphertext size, in bytes.  
\input{table/paramselect}

\subsection{Homomorphic Algorithms}\label{sec:background:algorithms}
Primitive HE operations presented in Table \ref{tab:ops} allow for SIMD
arithmetic to be performed on large vectors. These fundamental operations
compose to support computations such as convolution or matrix multiplication. 
Algorithms using HE primitives come in two varieties depending on how inputs are encoded into ciphertexts. {\em Batching} algorithms, optimized for throughput, encode a single data point from thousands of inputs into a single ciphertext vector and exploit natural SIMD operations\cite{CryptoNets, DNNTexas}. Alternately, {\em packed} algorithms, optimized for latency, encode multiple data points from a single input as a ciphertext vector and utilize permutations to properly align and operate on the desired points \cite{LoLa, Gazelle}.

\subsection{Homomorphic Applications}\label{sec:background:apps}
HE theoretically supports arbitrary functions (through mapping to polynomials
at extremely high computational cost \cite{CHET}), however,
practical HE applications use HE primitives directly and are
arithmetic-depth-limited.  Prior work has identified ML inference using deep
neural networks (DNNs) as a compelling use of HE, because of its structured form and complete reliance on simple linear algebra.  

Early work on HE for ML performed all computation on encrypted data \cite{CryptoNets, LoLa, DNNTexas}, communicating with the source of the data only to receive initial inputs and return final results. Unfortunately, these techniques have limited applicability to large and complex networks because of their modified activation functions, large noise growth across multiple layers, and subsequent reliance on very large parameter selections \cite{CHET}. This directly results in ciphertext sizes of multiple megabytes, inflating both computation and communication costs to impractical levels.

As an alternative, recent solutions have focused on {\em client-aided} HE \cite{Gazelle, nGraphHE2, Cheetah}. In these protocols, the natural linear algebra capabilities of HE are used to perform convolution and fully-connected layers on the offload server. The client is then enlisted to perform all non-linear activation and pooling operations. Intermediate results are communicated at layer boundaries and computation continues back-and-forth until the entire network is completed. 

Client-aided HE has been shown to boast multiple benefits. By avoid modified activation functions, they allow for privacy-preserving inference on pre-trained networks. Furthermore, by sending data back to the client for some operations, client-aided HE regularly refreshes the ciphertext noise budget. Thus, client-aided HE does {\em not} limit DNN depth,  and does {\em not} require extremely costly HE parameter settings. These benefits, however, come at the cost of increased client responsibility, especially for the abundant decryption and re-encryption operations. The client-side costs are the central impediment to the adoption of client-aided HE for IoT devices. \sys aims to break this participation barrier through encrypted algorithm optimization and architectural acceleration.

%% file: table/ops.tex
\begin{table}[!ht]
\renewcommand{\arraystretch}{1.3}
\caption{Homomorphic Encryption Operations: The operations available in homormorphic encryption along with their computational complexity and relative noise growth. All operations are performed with a ciphertext, i.e. a plaintext multiply denotes the multiplication of a plaintext with a ciphertext.}
\centering
\label{tab:ops}
\begin{tabular}{c|c|c}
\hline
Operation & Complexity & Noise Growth \\
\hline
Encrypt &  $O(N \times \log N \times r)$ & N/A\\
\hline
Decrypt &  $O(N \times \log N \times r)$ & N/A \\
\hline
Plaintext Add & $O(N \times r)$ & Small \\
\hline
Ciphertext Add & $O(N \times r)$ & Small \\
\hline
Plaintext Multiply & $O(N \times \log N \times r)$ & Moderate \\
\hline
Ciphertext Multiply & $O(N \times \log N \times r^2)$ & Large \\
\hline
Ciphertext Rotate & $O(N \times \log N \times r^2)$ & Small \\
\hline
\end{tabular}
\end{table}


%% file: table/params.tex
\begin{table}[htb]
\renewcommand{\arraystretch}{1.3}
\caption{Homomorphic Encryption Parameters for the BFV Scheme \cite{bfv1, bfv2, SEALmanual}}
\label{tab:params}
\centering
\begin{tabular}{c|l|l}
\hline
Parameter & Name & Description \\
\hline
$N$ & Poly. Mod. & \# of coeffs per ciphertext.\\
\hline
$q$ & Coeff. Mod. & Max value of ciphertext coeff. \\
\hline
$k$ & \# Coprime Mod. & Number of moduli in RNS \\
\hline
$\{k\}$ & Coprime Mod. Bits & Bits per coprime mod.\\
\hline
$w$ & Word Size & Bytes per encrypted coeff.\\
\hline
$t$ & Plaintext Mod.& Max value of plaintext coeff. \\
\hline
$s$ & Ciphertext Components & \# polynomials per ciphertext \\
\hline
\end{tabular}
\end{table}

%% file: table/paramselect.tex
\begin{table}[!ht]
\renewcommand{\arraystretch}{1.3}
\caption{HE Parameter Selections: All parameters are chosen to satisfy at least 128-bit security.}
\label{tab:paramselect}
\centering
\begin{tabular}{c|c|c|c|c|c}
\hline
Label & $N$ & $\log_2q$ & $\{k\}$ & $\log_2t$ & Size (B) \\
\hline
A & 8192 & 175 & \{58,58,59\} & 23 & 262,144\\
\hline
B & 4096 & 109 & \{36,36,37\} & 18 & 131,072\\
\hline
\end{tabular}
\end{table}


%% file: src/motivation.tex
\subsection{Motivation for Client-Aware Optimization}\label{sec:motivation}
Client-aided privacy-preserving computation places an enormous burden on resource-constrained IoT client devices. The computational costs intrinsic to the client-aided model motivate our work on client-optimized software and hardware support.

Interaction with the client is used to refresh the ciphertext noise budget and compute (relatively inexpensive) non-linear activation and pooling operations in plaintext. To do so, the system regularly exchanges data with the client to be decrypted, minimally computed on, and re-encrypted. This process has an important net effect on client responsibility: {\em abundant encryption and decryption operations are on the critical path}. These cryptographic operations in software are
extremely computationally costly, even using a highly-optimized commercially
available implementation~\cite{SEAL3.4}. 

We used our end-to-end software-only client-aided prototype to measure the cost of encryption and decryption in real encrypted neural network implementations. The software baseline system (which we describe in detail in Section \ref{sec:system}) utilizes existing server-optimized encrypted algorithms from \cite{Gazelle} and default parameter selections from \cite{SEAL3.4}.
We measure the time to complete a single classification inference using each
of four full-fledged DNN models (we describe our methodology in detail
in~\ref{sec:eval:method}).  

\begin{figure}[!ht]
\centering
\includegraphics[width=\linewidth]{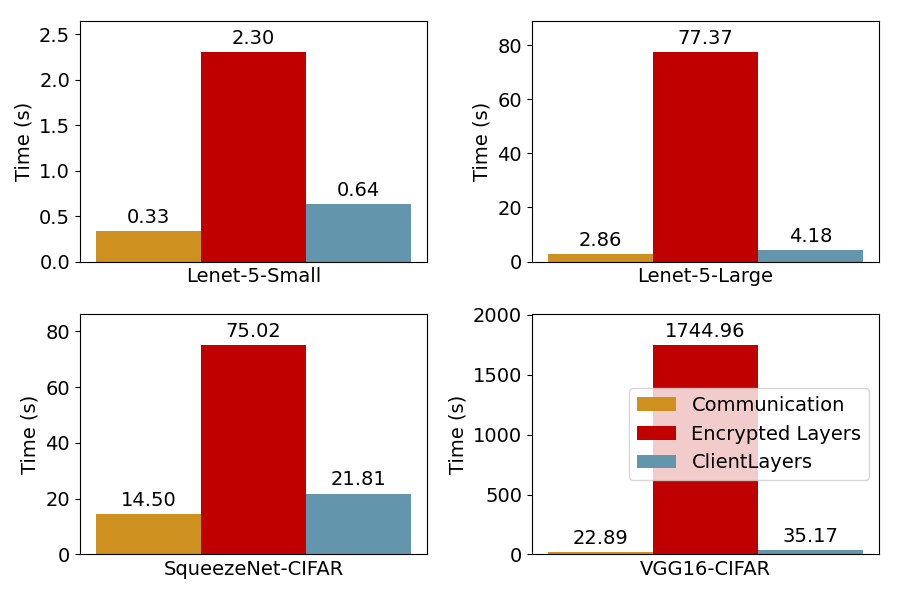}
\caption{Complete runtime in seconds for single image inference on each network.}
\label{fig:runtime}
\end{figure}

Figure~\ref{fig:runtime} shows the time spent computing on the client,
running encrypted computation on the offload, and communicating. As model
size increases, communication and client computation cost increases.
Encrypted computation cost at the offload device is consistently very time
consuming. A rich and complementary line of work in server-optimized encrypted algorithms \cite{LoLa, Gazelle, nGraphHE2, Cheetah, MiniONN, Delphi} and hardware support \cite{heax, HEAWS, heat} is expected to continue reducing these costs, but the goal of our work is {\em not} to optimize offload-side HE. Instead, we focus on reducing the
costs in time (and commensurately in energy) on the client side. Minimally addressed by prior work \cite{VLSIencrypt}, we approach this pursuit with specific attention toward realistic applications (e.g. parameters selections) and IoT devices (e.g. low-power). 

\begin{figure}[!ht]
\centering
\includegraphics[width=\linewidth]{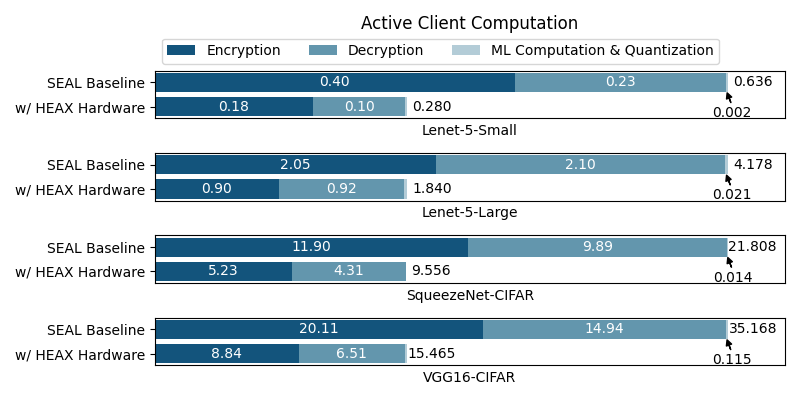}
\caption{Breakdown of client execution time in seconds using SEAL software with the default parameter settings on an unmodified ARM core and with hardware support for limited operations from \cite{heax}}
\label{fig:clientwntt}
\end{figure}

The compute costs on the client side are primarily the work of encryption and
decryption.  Figure~\ref{fig:clientwntt} shows a breakdown of time
spent by the client during these DNN inference computations.  Over 99\% of
the client's time is HE operations, rather than ML operations (i.e.,
non-linear computation \& quantization).  The plot also shows that existing
hardware support for the Number Theoretic Transform (NTT) and Dyadic
Multiplication, such as that provided in HEAX~\cite{heax}, is not sufficient to reduce the cost of encryption. We profiled the encryption and decryption computations in SEAL to determine that these operations only account for about 50\% of the total runtime. We optimistically modeled the benefit of hardware support for both sub-operations by scaling our software runtime accordingly by the speedup factor reported in the
original HEAX paper. Even with the the modular benefits of HEAX and other offload-side accelerators \cite{heax, HEAWS, heat}, encryption and
decryption time spent by the client remains dominant by orders of magnitude over ML computations.
Often dismissed as insignificant one-time costs, encryption and decryption are on the critical path in the client-aided model and demand further optimization via client-optimized encrypted algorithms and comprehensive hardware support. In this work, we develop such support, making client-aided encrypted inference feasible and favorable for resource-constrained IoT clients.

%% file: src/system.tex
\section{Client-Optimized Client-Aided HE}\label{sec:system}
\sys is a client-optimized system model and implementation for client-aided encrypted
computation.  The model assumes a resource-constrained client device and a
more computationally capable, but untrusted, shared offload server. Typical
of HE systems, we assume a semi-honest adversary model for the offload
device: the adversary may be curious about the input data, but the system is
trusted to faithfully perform the specified operations. 
In contrast to computationally expensive MPC protocols, \sys does not make any attempt to hide data on the offload device from the client, including pre-trained ML model data. Rather, the priority of \sys is to provide strong privacy guarantees for sensitive client data from IoT devices. 

\begin{figure}[!ht]
\centering
\includegraphics[width=\linewidth]{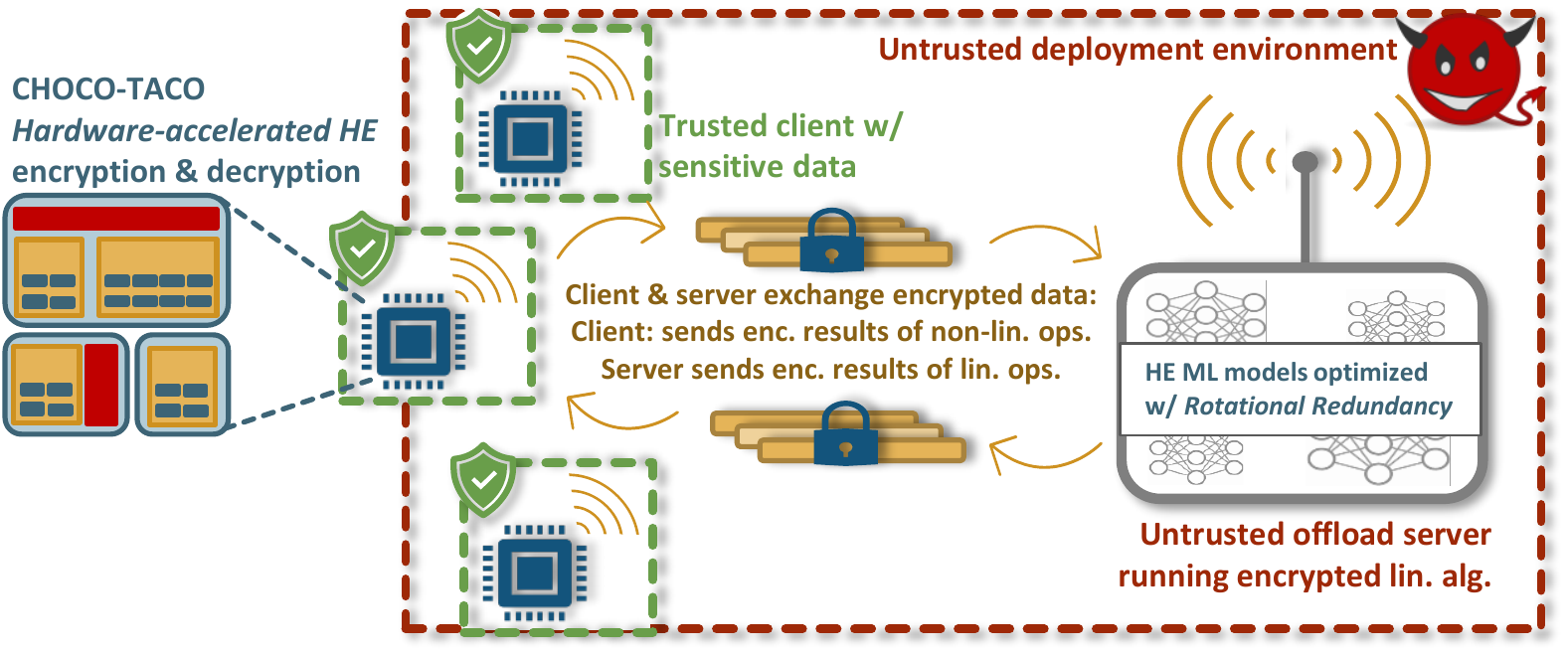}
\caption{System architecture for \sys. A resource-constrained sensor device and an untrusted offload device communicate via ciphertexts to collaboratively and securely process sensitive data.}
\label{fig:sysdiagram}
\end{figure}

\sys implements client-aided HE offloading, partitioning work layer-by-layer between the client and the offload device. It optimizes cryptographic
algorithms developed in prior work and tunes cryptographic parameters to
make feasible the amount of client work performed.  
\sys introduces {\em rotational redundancy}, which is a new approach to
permuting a vector of data encrypted in a cyphertext, which is useful for matrix manipulations such as convolution.  
The technique reduces
the noise growth imposed by common rotation operations, allowing for smaller parameter selections and correspondingly smaller ciphertexts. 

\subsection{Selecting Efficient HE Parameters}
\label{sec:system:params}
Choosing appropriate HE parameters (as introduced in Section~\ref{sec:background:HE}) is a vital yet cumbersome step of encrypted application development \cite{CHET}. The selection of different parameters leads to different ciphertext size and noise characteristics, which, in turn, influence computation, encryption, decryption, and communication costs in the client-aided model.
A system can achieve the same security level with different parameters (e.g.
different ciphertext sizes). As such, \sys actively minimizes parameter selection via quantization and encrypted algorithm optimization.
%

An HE scheme's plaintext modulus $t$ defines the number of bits in which to store
each plaintext value within each element of an encrypted vector.
\sys quantizes data into fewer bits before encrypting, because
when tolerable, quantization to fewer bits allows the use of a smaller plaintext modulus. As shown by the varying $t$ values in Table \ref{tab:rotatenoise}, this increases a ciphertext's noise budget without changing its size, incentivizing the smallest possible $t$ that does
not overflow~\cite{NNDistiller}.  Our \sys prototype quantizes signed
floating point input values to a 4-bit range. The sequentially applied HE
operations of convolution then expand these values up to (but not over) our
23-bit prime coefficient modulus value. When a ciphertext is refreshed at layer boundaries the data is also requantized to 4-bits.

\subsection{HE Algorithm Optimization}
\label{sec:system:algos} 
\sys uses the cryptographic primitives from the SEAL encrypted computing
library~\cite{SEAL3.4} to implement the HE algorithms developed in
Gazelle~\cite{Gazelle}. These HE algorithms correspond to encrypted linear
algebra operations used, for instance, in DNN inference.  In \sys, all of
these encrypted computations execute on the offload device.

Introduced in Section~\ref{sec:background:algorithms}, Gazelle implements {\em packed} homomorphic algorithms which require permutations for rearranging input elements within an encrypted vector. In this way, elements are appropriately aligned for matrix manipulations such as 1-D, 2-D, and strided
convolutions~\cite{Gazelle}.
A key challenge presented by arbitrary permutations is that each requires a
sequence of encrypted vector rotation and masking
multiplication operations~\cite{HELibAlgs}.
The depth of such operations quickly deplete the limited noise budget of a ciphertext. 

\begin{figure}[!ht] \centering
\includegraphics[width=0.9\linewidth]{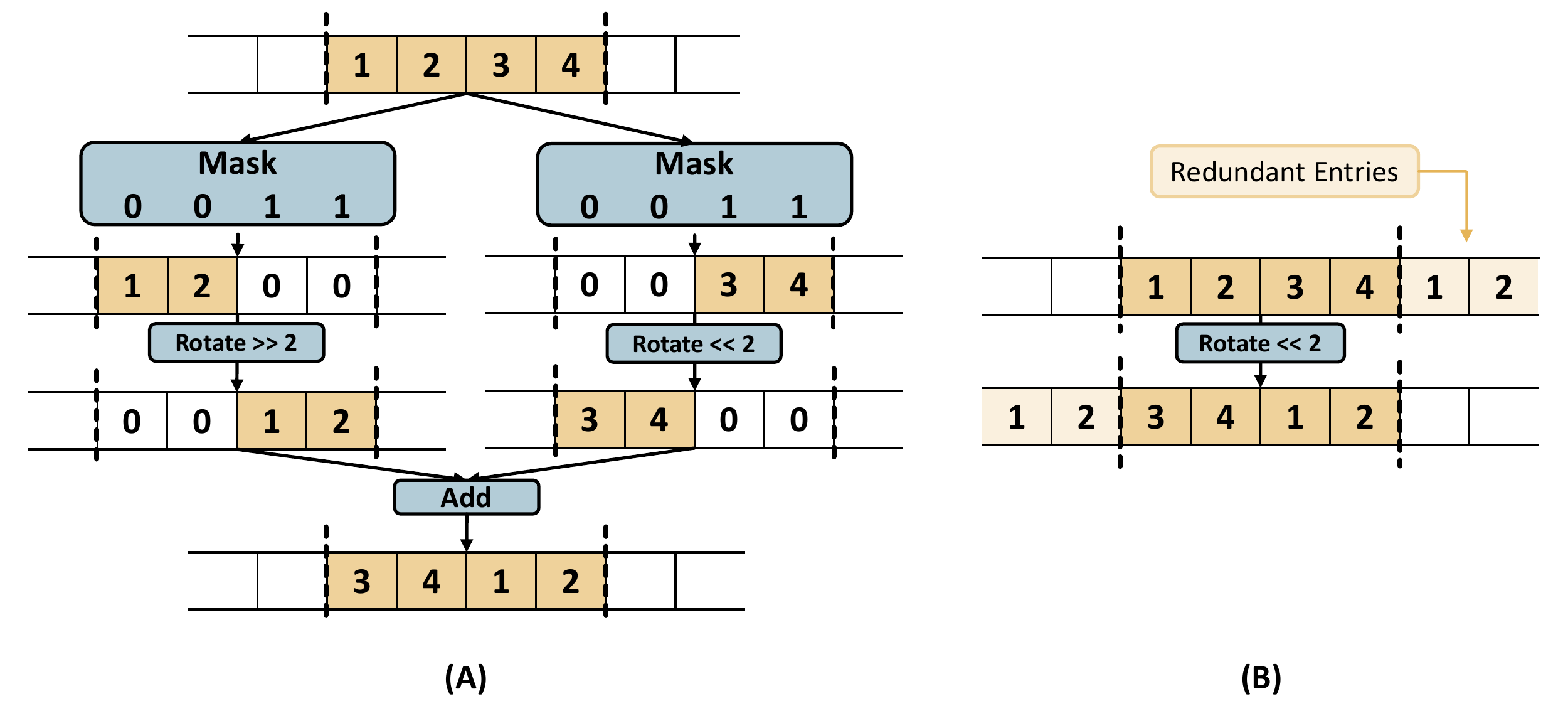} 
\caption{Encrypted windowed rotation using arbitrary permutation (A) and rotational redundancy (B)}
\label{fig:rotationalgs} 
\end{figure} 

Rotational redundancy is a new way to perform certain permutations on an
encrypted vector. The technique targets {\em windowed rotation} permutations
that rotate the elements in a sub-range of a vector, wrapping elements around from the top of the sub-range to the
bottom and vice versa. This is in contrast to naturally supported HE rotations which can only rotate the vector in its entirety. 
%
Figure~\ref{fig:rotationalgs} shows a windowed rotation permutation on a ciphertext. The standard implementation (A) uses
both rotations and masking multiplications, quickly exhausting the ciphertext's
noise.
In contrast, implementation (B), introduced in \sys, uses rotational redundancy to perform such a permutation with only a single relatively low-cost encrypted
rotation.
The key to rotational redundancy is to pack the window of values to be
rotated with additional redundancy on either side {\em before} encryption. The redundancy contains the in-window elements that would
``wrap around'' when the permutation executes.  After a series of windowed rotations and other
operations, values outside the window of interest are simply ignored
upon decryption.  
The amount of redundancy required for a windowed rotation corresponds to the amount of rotation to be performed, and, for the rotations used in convolutions, is typically a small fraction of the vector size. 
Rotational redundancy trades the use of more space in a vector for slower
depletion of ciphertext noise, and in turn enables the use of smaller parameter selections.
Table~\ref{tab:rotatenoise} quantifies these benefits.

\input{table/rotatenoise2} 

In our neural network image classification implementation, a ciphertext
vector is the concatenation of a vector per channel in the image.  Executing
inference requires windowed rotation within each channel. We pack images,
adding rotational redundancy to each channel.
By then packing channels into evenly spaced power-of-two slots in the ciphertext, the alignment of entire channels can also be achieved with simple encrypted rotations and no masking multiplies. Ultimately, convolution is achieved with optimal multiplication efficiency - a single multiplication of the weights with the inputs.

 After all algorithm and parameter optimizations, \sys utilizes freshly encrypted ciphertexts with only 2 prime residues, a 50\% reduction in ciphertext size from the SEAL default parameters for $N=8192$. Half of that improvement, the elimination of one entire residue, is from rotational redundancy alone. As discussed in Section \ref{sec:evaluation}, this reduction in ciphertext size has a direct and dramatic benefit in both computation and communication costs.

\par

%% file: table/rotatenoise2.tex
\begin{table}[!ht]
\renewcommand{\arraystretch}{1.3}
\caption{Noise Budget: The initial noise budget of a ciphertext varies with different selections of $N$, $\log_2t$, and $\{k\}$. The noise budget remaining after a single rotation versus an arbitrary permutation with masking also demonstrates the benefit of rotational redundancy in eliminating masking multiplies.}
\label{tab:rotatenoise}
\centering
\begin{tabular}{c|c|c|c}
\hline
Parameters & Initial & After Rotate & After Permute\\
\hline
8192, 20, \{58,58,59\} & 68 & 66 & 42\\
\hline
8192, 23, \{58,58,59\} & 62 & 59 & 33\\
\hline
8192, 28, \{58,58,59\} & 52 & 50 & 18\\
\hline
4096, 16, \{36,36,37\} & 33 & 31 & 12\\
\hline
4096, 18, \{36,36,37\} & 29 & 26 & 5\\
\hline
4096, 20, \{36,36,37\} & 25 & 22 & 0\\
\hline
\end{tabular}
\end{table}

%% file: src/architecture.tex
\section{Hardware Acceleration}
\label{sec:arch}
\arch is a hardware accelerator for homomorphic encryption and decryption
operations, designed for client-aided HE.
Figure \ref{fig:clientwntt}, shows that accelerating NTT/INTT and dyadic
multiplication~\cite{heax} only is insufficient in reducing the dominating costs of these cryptographic primitives.  \arch, by contrast,
effectively accelerates all of the component functions that make up HE
encryption and decryption, which we demonstrate through a worked example in
Section~\ref{sec:arch:example} and show quantitatively in
Section~\ref{sec:designspace}.

\subsection{BFV Encryption}\label{sec:arch:encryption}
\begin{figure}[!ht]
\centering
\includegraphics[width=\linewidth]{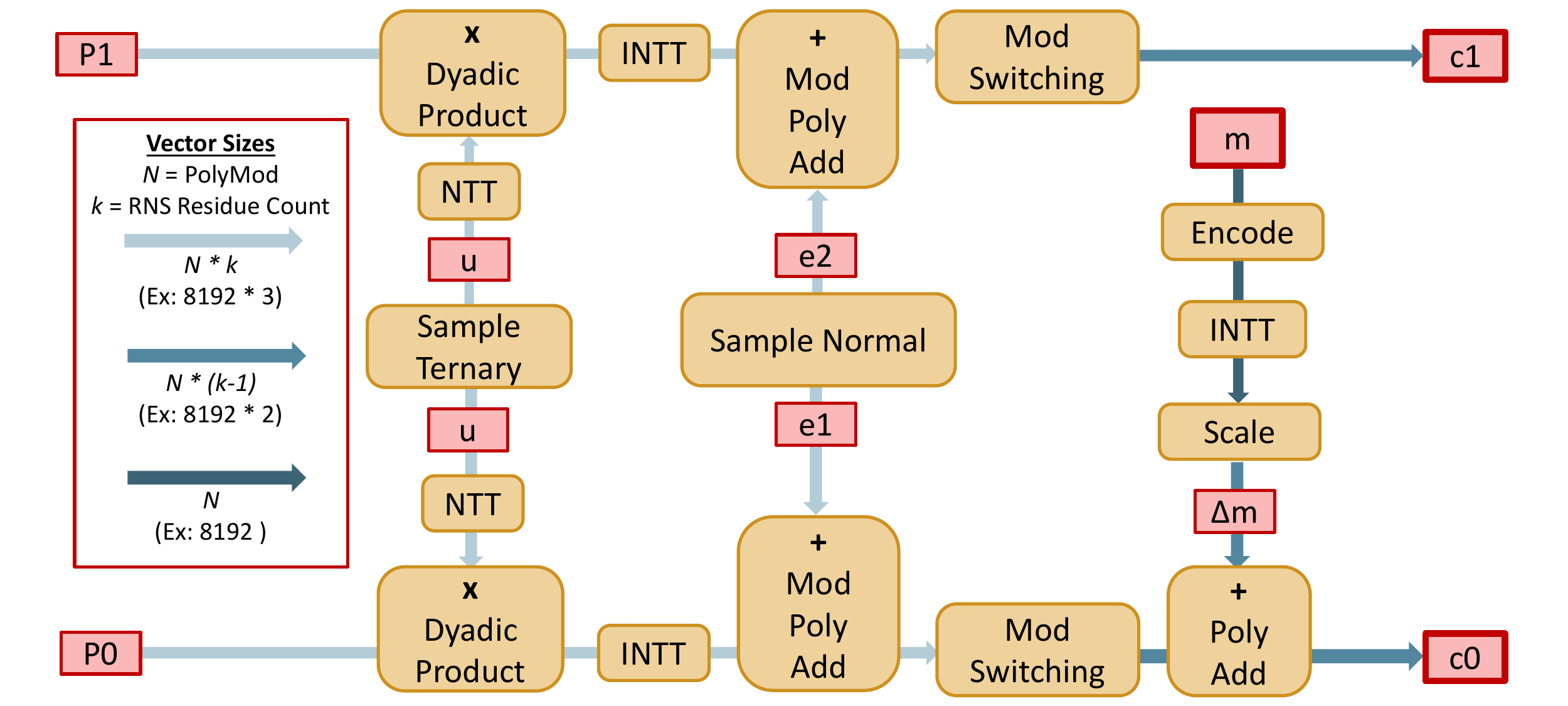}
\caption{Pipeline of the BFV encryption operation to sample random noise, scale the input message, and ultimately create a ciphertext of two polynomials, each in RNS form. \cite{SEALmanual, HEAWS}}
\label{fig:encryptpipeline}
\end{figure}

{\small
\begin{equation}\label{eq:encryption}
\begin{split}
Enc([P0, P1], m) = ([\Delta m + P0u + e1]_q , [P1u + e2]_q) \\
where \ u \stackrel{\$}\leftarrow \ R_2 \ and \ e1, e2 \leftarrow \chi \quad \quad \quad \quad \quad
\end{split}
\end{equation}
}
\arch accelerates asymmetric BFV encryption, described by Equation
\ref{eq:encryption}~\cite{SEALmanual}, where $m$ is a message to
encrypt, $P0, P1$ are public keys, and $u, e1, e2$ are vectors of randomly sampled numbers. 
Figure~\ref{fig:encryptpipeline} fully diagrams the RNS implementation of
equation~\ref{eq:encryption} from SEAL~\cite{SEAL3.4}. 
The algorithm encrypts a message by first producing an encrypted "zero" by combining the vectors of randomly sampled numbers with the public keys through coefficient-wise multiplication and addition. The encoded message is then added to the encrypted zero to produce the final ciphertext.

\begin{figure*}[t]
\centering
\includegraphics[width=\linewidth]{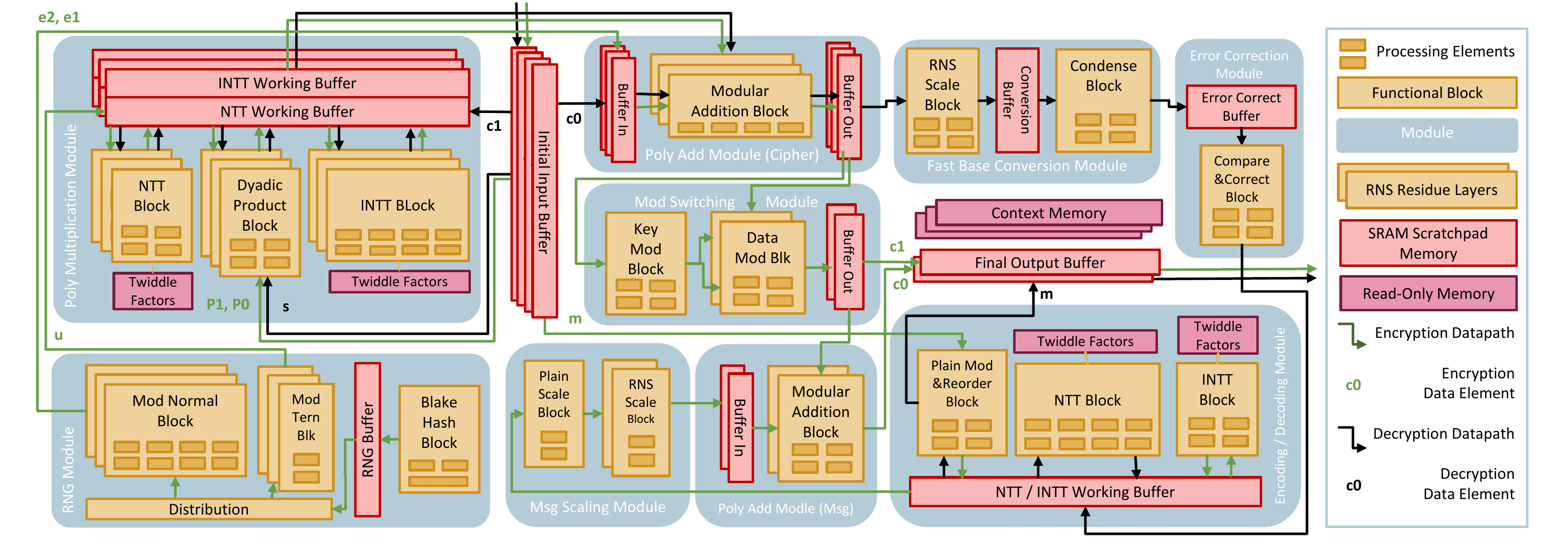}
\caption{Specialized architecture to support encryption \& decryption for $k = 3$}
\label{fig:encryptarch}
\end{figure*}

\subsection{Encryption Architecture}
\label{sec:arch:arch}
\arch accelerates each of the main sub-operations that make up BFV encryption
and decryption.  \arch is a parallel, pipelined accelerator that directly
implements the BFV algorithm.  Figure \ref{fig:encryptarch} shows the full
encryption and decryption accelerator.
The design has several key {\em modules}: Random Number Generation,
Polynomial Multiplication, Polynomial Addition and Modulus Switching, and
Message Encoding. Each module contains functional blocks, which may be
replicated for parallelism, and memory necessary to perform each operation.
Each functional block contains an array of processing elements to allow for
data parallelism within a functional block.  The design additionally pipelines operations within and across modules.  \par

\subsubsection{Random Number Generation}
\arch has a dedicated RNG module that implements the Blake3~\cite{Blake3}
cryptographic hashing algorithm.  The \arch configuration in Figure
\ref{fig:encryptarch}, requries this module to produce 565 MB/s of random values
at peak and 201 MB/s on average.
The RNG module is also responsible for interpreting the provided randomness as an
integer from either a ternary or normal distribution, which it transforms
into RNS form.
We upgraded our software encryption implementation to use
Blake3 instead of Blake2, giving the algorithmic performance
increases to both \arch and the baseline.  \par

\subsubsection{Polynomial Multiplication}
\arch includes a module for polynomial multiplication of two polynomials, like
prior work~\cite{heax, HEAWS, VLSIencrypt, heat}. The module transforms the polynomials to NTT form, then
performs element-wise dyadic product.  The module converts the result back to a
polynomial using INTT.  \arch's NTT and INTT modules are conceptually similar
to HEAX~\cite{heax}.  The modules perform pipelined SIMD memory accesses,
following the NTT's butterfly dataflow pattern.

The hardware computes polynomial multiplication of 
the RNS form of $u$ and each public key, $P0$ and $P1$. 
SEAL stores public keys in NTT form, so only the single NTT transformation for $u$ is supported in hardware.
NTT's butterfly dataflow requires access to an entire polynomial at once,
precluding aggressive pipelining (i.e., forwarding partial results).
Once $u$ is entirely transformed, results flow to the dyadic product block along with $P0$, and $P1$.
Both ciphertext components ($c0$ and $c1$) use the same NTT encoding of $u$,
allowing it to remain in the NTT working buffer throughout encryption.
Dyadic multiply results flow to a separate INTT buffer.  
After all outputs accumulate, the hardware performs in-place INTT, producing
a result. \par

\subsubsection{Polynomial Addition and Modulus Switching}
After polynomial multiplication the accelerator applies polynomial addition
and modulus switching. 
Polynomial addition is coefficient-wise addition of polynomials, taking one
input from the INTT buffer and the other from a buffer filled by the RNG
module. 
The output fills a small intermediate buffer that is the input to modulus
switching, which removes the key-prime residue from the RNS encoding, resulting in $k-1$ polynomials. 
Modulus switching applies a series of modular multiplications and reductions,
Modulus switching is the only operation that requires interaction across RNS residues, which precludes straightforward data parallelism across residues. 
\par


\subsubsection{Message Encoding}
After encrypting zero, \arch encodes, scales, and adds the input
message to the encrypted zero. 
The encode/decode module includes a pair of small NTT and INTT blocks.  
The encoding hardware computes the modulus of each coefficient by the
plaintext modulus $t$, and reorders coefficients into slots of the plaintext. 
The hardware must convert the encoded message to RNS, by scaling,
representing it with $k-1$ residues only.  A dedicated polynomial addition
module adds the result to the $k-1$ intermediate residues of $c0$ and stores the result in the final output buffer.

\subsubsection{Memory}
Each module integrates and manages embedded SRAM scratchpad memory.
All modules except NTT accept streaming inputs, and a module's memories must
accommodate its incoming (parallel) input stream at the rate of input
arrivals and operation duration.
NTT and INTT, however, algorithmically operate on an entire polynomial,
requiring their buffers to be sized according to the HE scheme's polynomial
size.
With HE parameters $N=8192$ and $k=3$, each NTT/INTT buffer is 64kB. 
In contrast, other memories, sized via our design-space exploration in
Section \ref{sec:designspace} are sub-1kB. 
As Section \ref{sec:designspace} explains, we model memories using Destiny,
ported for single-reader, single-writer 64-byte data accesses.

\subsection{Encryption Operation Example}
\label{sec:arch:example}
BFV encryption encrypts zero into a ciphertext and adds a scaled message to
the encrypted-zero ciphertext. 
To start, the accelerator samples $N$ bytes from the RNG according to a ternary distribution storing them in the NTT working buffer as $u$.
The NTT block produces the NTT of $u$ in place, and the value becomes the
input to the dyadic product module. 
The dyadic product module's other input is the accelerator input buffer,
which software initializes with the NTT-transformed residues of $P1$, a 
public key.
Dyadic multiply produces the element-wise product of $u$ and $P1$ in the INTT
block's buffer, which the INTT block processes.

In parallel with the dyadic product of $u$ and $P1$ the RNG unit produces $e2$, a
sequence of 8-byte, normally distributed samples, storing them in a buffer in
the cipher addition module. 
When the INTT completes, the PolyAdd module streams in its result, performing
element-wise, addition with $e2$ and storing the results in a buffer in the
modulus switching module, as $c1$.\par
$c1$ is one component of the final ciphertext, which is output into the CPU's
host memory

Meanwhile, the accelerator begins producing $c0$ in the Poly Multiplication
module.  
Computing $c0$ reuses the NTT of $u$, performing element-wise
multiplication with public key $P0$.
The accelerator samples a sequence of normally-distributed 8-byte $e1$ values
(like $e2$) adding them element-wise with the result of the INTT of  the
product of $u$ and $P0$.  The result is a partially-computed version of
$c0$, the other component of the ciphertext.\par
The last step is the polynomial addition of the partially computed $c0$ and
the encoded input message, producing $c0$, which together with $c1$ makes the
final ciphertext.
%
%
\subsection{Adding Parallelism} \label{sec:arch:designspace} 
\arch exploits pipeline and data parallelism
available in BFV encryption.  Parallelism exists in independent RNS residues,
independent coefficients and in pipelining throughout the accelerator. 

Polynomial multiplication and addition manipulate the multiple residues of an
RNS-encoded polynomial.  
Each of these residues is an independent share of the polynomial being
manipulated and operations need to be applied identically to each residue.
Up to the limits of area and power, a \arch architecture can create parallel
replicas of these operations' modules, including their input and output
memories, enabling parallel processing of a polynomials RNS residues. RNS parallelism  also eliminates the need to buffer the large vectors of random number for future execution. Instead, $u, e1,$ and $e2$ are immediately consumed and distributed to all residues as they are generated.
Figure \ref{fig:encryptarch} illustrates the parallelism of RNS residue
operations graphically through layering.

Within each RNS layer, thousands of coefficients per polynomial afford data
parallelism.
A key design parameter for the \arch architecture is the degree to which each module exploits this source of data parallelism. 
Up to the limits of area and power, a \arch architecture can create parallel
replicas of the blocks in a module, sizing memories to match, to enable
higher throughput processing of coefficients. 
Section~\ref{sec:designspace} systematically explores the design space of
parallel accelerators.

\subsection{Decryption Support}\label{sec:arch:decryption}
BFV decryption is operationally very similar to encryption. Equation
\ref{eq:decryption} shows decryption mathematically. 
\begin{equation}\label{eq:decryption}
Dec(s, [c0, c1]) = \left[\bigg\lfloor\frac{t}{q}[c0 + c1s]_q\bigg\rceil\right]_t
\end{equation}

Figure~\ref{fig:encryptarch} shows the flow of control and data for
decryption with black lines.  Decryption requires a few additional hardware
components, but reuses the existing polynomial
multiplication and addition modules to process $c1$, $s$, and $c0$. 
After addition, these intermediate results
undergo fast base conversion and error correction, after which
the message need only be decoded. 
Decoding uses the message encoding module, performing  
NTT, then moding by the plaintext modulus $t$.
The result is the decrypted message, which the hardware conveys to the CPU's
memory.



%% file: src/designspace.tex
\section{Architectural Design Space Exploration}\label{sec:designspace}
We explore the design space of the \arch hardware using a custom simulation
infrastructure.  The hardware model captures the effects of parallelism and
pipelining and estimates time, power, area, and energy.
We implemented individual hardware components in RTL and synthesized them with
Cadence Genus, in a generic 45nm technology node. We modeled three-stage,
pipelined multiplication and division units.
To model memory, we used Destiny~\cite{destiny}, modeling SRAMs using its
aggressive wire technology, optimized for read energy with 8-word, 64-byte,
memory accesses.
The access latency of our energy-optimized memories limits clock frequency, 
and we clocked the design at 100 MHz.

\subsection{Performance Tradeoffs}
\begin{figure}[!ht]
\centering
\includegraphics[width=\linewidth]{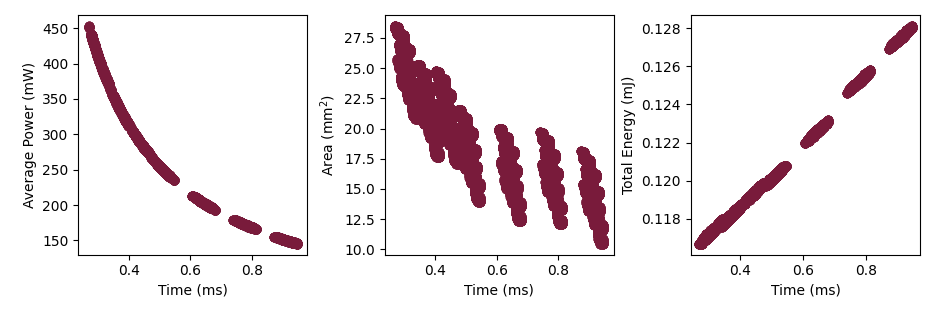}
\caption{Design space for encryption hardware with respect to power, area, and energy. Parallelism tradeoffs are available in multiple dimensions at each stage of the encryption pipeline.}
\label{fig:designspace}
\end{figure}

We quantified the tradeoff of area, time, and power, with a systematic
exploration of the \arch hardware design space. 
Using our simulator, we swept across 31,340 different architectural configurations.
For each block in each module, the study varied the number of processing
elements from one to 16 in powers of two, varying memory capacity
commensurately, between 128 and 1024 bytes per RNS-parallel layer, except for
NTT/INTT units, which require a fixed memory size.
For each configuration, the study assessed power (leakage \& average
dynamic), area, energy and compute time for a single encryption operation.
Results from the design exploration are presented in Figure
\ref{fig:designspace}. \par

%
Overall, the design space shows significant variation in power and area, with
a marked Pareto frontier along which power, time, and area balance.  We
selected an operating point for \arch by limiting power to 200 mW, and
choosing the smallest design that had a run time within 1\% of the best run
time (and energy). 
The chosen configuration has \totalarea area and consumes \encenergy to perform
a single encryption in \enctime. Figure \ref{fig:encryptarch} depicts this
configuration graphically.

\subsection{\arch Makes Encryption Fast and Low-Energy}\label{sec:designspace:scalability}
\begin{figure}[!ht]
\centering
\includegraphics[width=\linewidth]{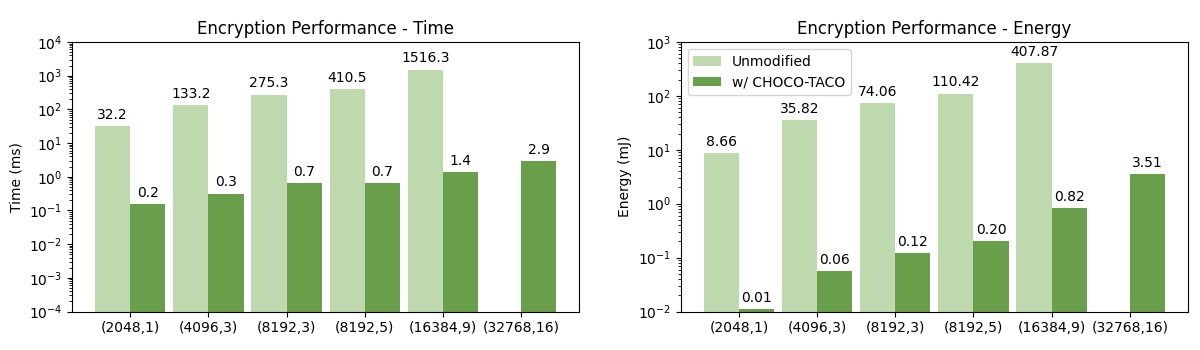}
\caption{Logarithmic comparison of time \& energy of the \arch architecture presented in Figure \ref{fig:encryptarch} for varying encryption parameters versus a 528MHz device w/o dedicated hardware support.}
\label{fig:scalability}
\end{figure}

We evaluated the benefit in time and energy of \arch for encryption compared
to a software encryption baseline, showing that across a range of HE
parameter settings, hardware support provides substantial improvements.
Figure~\ref{fig:scalability} shows data comparing software encryption to
encryption with hardware acceleration.
%
%
We evaluated the default HE parameter settings of SEAL, as well as \sys's parameter setting of (8192, 3), as presented in Table \ref{tab:paramselect}.  
The baseline is
an average of 100 encryption operations running in software on our 
IMX6 hardware. Results are shown in Figure
\ref{fig:scalability}. We omit baseline data for the (32768,16) parameter
setting because the IMX6 board does not have enough memory to
encrypt data for these parameters. Notably, this configuration with its prohibitive memory requirements is not uncommon in existing encrypted inference solutions \cite{CHET, LoLa}

For the \sys (8192, 3) configuration, \arch provides an improvement over the software baseline of \encspeedup in time and \encenergysave in energy. 
The data also show a performance scaling trend, in that with hardware
support, encryption time scales up directly with $N$, while software scales
up with both $N$ and $k$.
The scalability benefits comes from parallelism in the accelerator
architecture: replicated modules process independent RNS residues in parallel.
Decryption sees less benefit from hardware acceleration than encryption, with
only a \decspeedup speedup over software for the (8192,3) \sys parameter
selection. This decrease in speedup is contributed to limited parallelism because decryption operates on only a single polynomial at this parameter selection.

Overall, \arch provides up to \encmaxspeedup and \encmaxenergysave savings in
time and energy, respectively, and providing consistent gains across HE
parameter settings.\par

%% file: src/evaluation.tex
\section{Evaluation}\label{sec:evaluation}
We evaluate \sys to show the importance of algorithm optimization and architectural acceleration to the practicality of resource-constrained devices participating in client-aided privacy preserving computation. In accordance with prior work, we evaluate \sys using several large-scale machine learning models. We show reduction of communication overhead by up to three orders of magnitude over existing state of the art privacy-preserving DNN inference solutions. This is a direct result of the smaller parameter selections enabled by \sys optimizations, including rotational redundancy. Additionally, we show that the comprehensive hardware acceleration provided by \arch improves the runtime of active client computation by an additional \tacovsheax than dedicated NTT/INTT and dyadic product acceleration alone. Ultimately, we demonstrate that a complete \sys implementation using Bluetooth communication is comparable to local computation with TensrFlow Lite and, for large networks, reduces inference energy by up to 37\%.

\subsection{Applications \& Methodology}\label{sec:eval:method}
\subsubsection{Neural Network Selection}\label{sec:method:nn}
We evaluate \sys for DNN inference, implementing client-aided, encrypted versions of the four image classification DNNs in Table \ref{tab:networks}. 
The Lenet variants operate on MNIST data~\cite{MNIST}, and the other, larger
networks classify CIFAR-10 images~\cite{CIFAR10}. 
We trained the DNNs on unencrypted data
using standard quantization-aware training in Tensorflow 2.2.0~\cite{Tensorflow2.2}. Evaluations are performed by running single-image inference through each network.

\input{table/networks}

\subsubsection{Client Modeling}\label{sec:method:system}
We perform baseline client evaluations for software running on an NXP IMX6
evaluation kit with an ARM Cortex-A7 CPU at 528MHz, 32/128 kB of L1/L2 cache,
and 4 GB DDR3L SDRAM.  We estimate power and energy using an average power
characterization (running Dhrystone) of 269.5 mW from the manufacturer's
Application Note AN5345~\cite{imx6appnote}. We follow the methodology from Section \ref{sec:background} to estimate HEAX acceleration and use the hardware configuration from Section \ref{sec:designspace} to model \arch acceleration.
\subsection{\sys Optimizations Reduce Communication}\label{sec:eval:mpccomm}
The algorithmic optimizations presented in Section \ref{sec:system:algos}, including rotational redundancy, minimize noise growth to enable smaller parameter selections and correspondingly smaller ciphertexts. As discussed in Section \ref{sec:background:HE}, all of the networks included in Table \ref{tab:networks} can be evaluated in \sys using ciphertexts with no more than 8192 elements ($N=8192$). This is in contrast to existing HE solutions \cite{Gazelle, CHET, Cheetah, CryptoNets, LoLa, MiniONN} which commonly use ciphertexts of 16 or even 32 thousand elements. By eliminating unnecessary prime residues, \sys further reduces ciphertext size by another 50\% over SEAL's default parameters at $N=8192$. 

\begin{figure}[!ht]
\centering
\includegraphics[width=\linewidth]{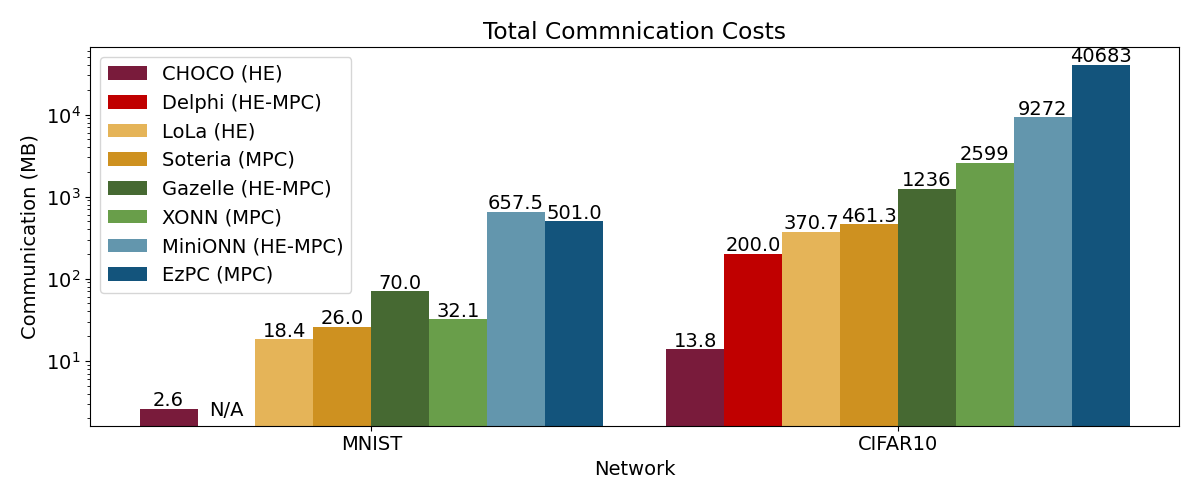}
\caption{Comparison of total communication requirements to perform single image inference via the Lenet-5-Large (MNIST) and SqueezeNet (CIFAR-10) implementations in \sys and comparable networks in several state of the art privacy-preserving DNN protocols.}
\label{fig:mpccomm}
\end{figure}

\begin{figure}[!ht]
\centering
\includegraphics[width=\linewidth]{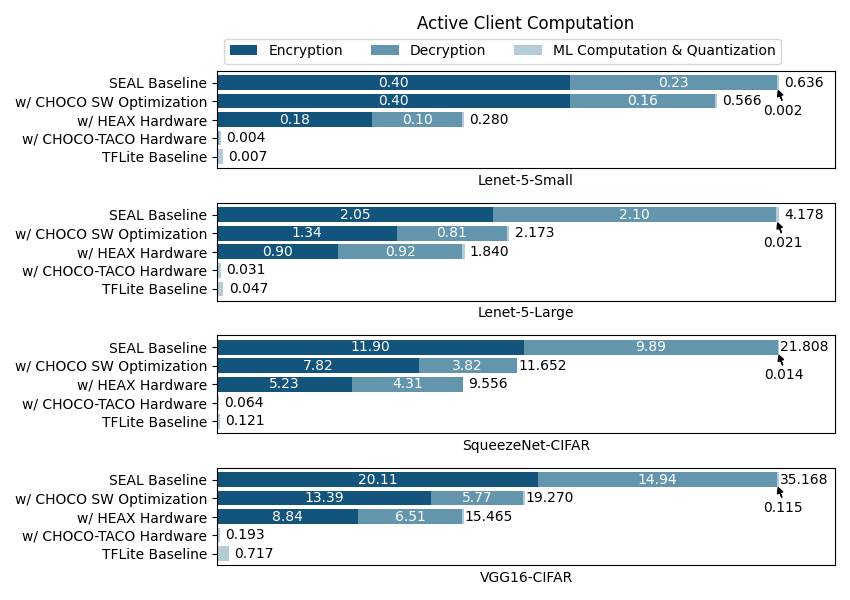}
\caption{Extension of client execution times shown in Figure \ref{fig:clientwntt} including \sys execution time breakdown.}
\label{fig:clientwchoco}
\end{figure}

This reduction in ciphertext size directly translates to improvements in server runtime, client responsibility (Figure \ref{fig:clientwchoco}), and communication overhead. Figure \ref{fig:mpccomm} demonstrates the communication improvement over several state of the art privacy-preserving DNN protocols. Comparisons are evaluated between the \sys implementations of Lenet-5-Large and SqueezeNet and comparable networks performing MNIST and CIFAR-10 single-image inference, respectively. They include communication for both offline preprocessing and online computation. Although the networks evaluated in this work are substantially larger (more model parameters) in all cases, \sys outperforms existing protocols by up to three orders of magnitude. Because of the smaller parameter selections and more efficient ciphertext packing, benefits are witnessed even compared to LoLa\cite{LoLa}, a \emph{non}-client-aided encrypted inference protocol. For the most closely comparable protocol, namely Gazelle \cite{Gazelle}, \sys still provides a nearly $90\times$ improvement in communication overhead. This reduction dramatically reduces end-to-end latency, especially for IoT devices often communicating over lower bandwidth channels such as Bluetooth.

\subsection{\arch Accelerates Client Computation}
We evaluate our hardware setup running single-image inference. We compare it to both the software optimized baseline and the baseline equipped with HEAX's NTT
unit and Dyadic Multiply unit only. The software optimized baseline includes the algorithmic optimizations of \sys, namely rotational redundancy, and already demonstrates an average \avgswspeedup improvement over the SEAL baseline software with standard permutations and default parameter selections. A baseline that runs local inference on the ARM Cortex-A7 CPU using TensorFlow Lite (TFLite) software is also included as a lower-bound.
We compute time and energy savings for active client computation by counting
the number of encryption and decryption operations necessary to run inference
for each network, and multiplying by the cost of each operation.
We assume that the time for client computation and
quantization, including ReLu and Pooling, stays the same for \sys as in
the baseline. 
Figure \ref{fig:clientwchoco} reports the resulting execution times, breaking down
the total time into its constituent components.
The data show an average speedup of \avgnnspeedup over the optimized software baseline for computations on the
client, which is consistent with the \encspeedup and \decspeedup speedups
observed for encryption and decryption, respectively. 

The data clearly show that encryption and decryption are the bottleneck on the client. NTT and Dyadic multiply only account for roughly 50\% of these operations. Thus, dedicated hardware units for these sub-operations alone, including but not limited to those provided by HEAX\cite{heax}, is not enough. The cryptographic operations for a client-aided protocol are still \avgheaxtimeovhd slower on average than computing the entire network locally with TF Lite. 

Comprehensive hardware acceleration for encryption and decryption is imperative. \arch recognizes this and uses an optimal allocation of compute resources, minimal buffering, tightly integrated memories, and multiple levels of parallelism to address the remaining 50\% of computation. With the acceleration provided by \arch, the time of active client computation on client-aided encrypted DNN inference becomes \avghwvstflite faster on average than local inference.

\subsection{Full Implementation Comparable to Local Compute}
\begin{figure}[!ht]
\centering
\includegraphics[width=\linewidth]{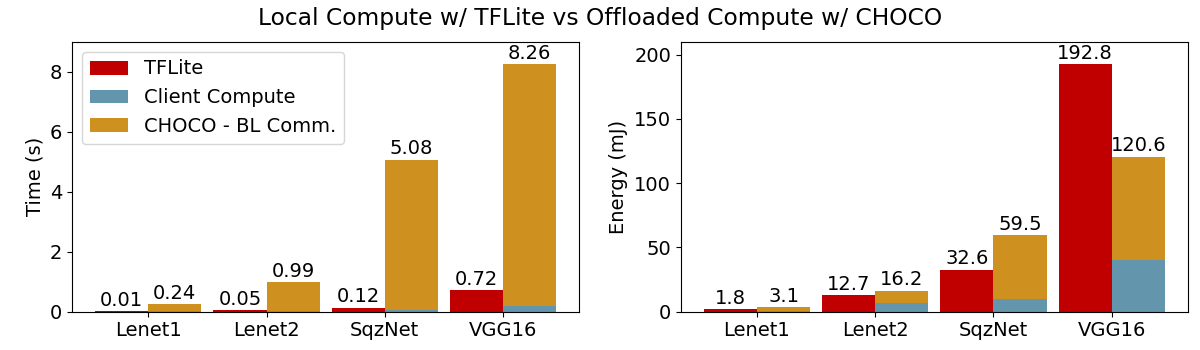}
\caption{Client execution time \& energy for single image inference via local compute using TFLite software and offloaded compute using a full hardware-software reference implementation of \arch using Bluetooth communication.}
\label{fig:vstflite}
\end{figure}

To understand the end-to-end benefits, we study a reference implementation of \sys that
communicates between the client and the offload using 10mW Bluetooth communication at 22 Mbps\cite{MECcomm}. Timing and energy results follow analytically from the data communication requirements, included in Table \ref{tab:networks}, of each network. End-to-end time and energy results are compared against the TFLite baseline in Figure \ref{fig:vstflite}.
The data show that in a full implementation communication time begins to dominate. For low-power, low-data-rate protocols such as Bluetooth, communication presents a \blavgoverhead average time overhead compared to local compute. However, in small devices battery preservation often outweighs the need for fast compute. In energy consumption \sys is competitive with the TFLite baseline. For VGG, the largest and most complex of the DNNs evaluated, \sys
earns up to a \vggenergysave end-to-end decrease in energy consumption.

The data carry several main take-away points.
The first take-away is that
hardware acceleration, like \arch is essential to make feasible the \sys
model for client-aided encrypted computing.  Without our hardware
acceleration --- even with the partial acceleration from HEAX~\cite{heax} ---
encryption and decryption are the computing and energy bottleneck.  Our
hardware support accelerates the entire encryption and decryption
computation, driving its cost down, eliminating it as the bottleneck, and
making \sys feasible. 
Second, intentional client-aware optimizations are essential to bring privacy-preserving computation to resource-constrained IoT clients. Although communication remains a key bottleneck in time and energy, the algorithmic optimizations of \sys reduce this cost by up to three orders of magnitude. For the first time, this dramatic reduction makes client time and energy requirements competitive with local inference, even displaying the possibility of end-to-end gains.
Third, the benefit of \arch depends on the structure of the
computation: VGG sees substantial performance and energy improvements, while
SqueezeNet sees a break-even or loss.  We characterize this workload-dependent benefit in the next section.

\subsection{Network Design}
\begin{figure}[!ht]
\centering
\includegraphics[width=0.95\linewidth]{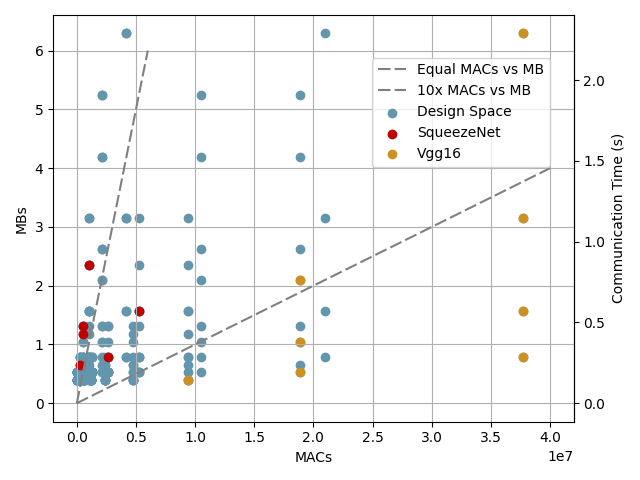}
\caption{Communication vs Computation for Convolution Layers of a DNN with different parameters}
\label{fig:layerstudy}
\end{figure}

Different workloads see different benefit from \sys computation, which owes
to differences in the rates of computation and communication required by
these different workloads. 
\sys using Bluetooth communication for VGG sees a \vggenergysave
energy savings, while SqueezeNet sees a \sqzenergyoverhead energy overhead. 
We performed a microbenchmarking study to evaluate this influence of workload
structure.
We constructed workloads with a variety of different convolutional DNN layers
of different dimensions.  The structure of the convolutional layers varies
the number of multiply-accumulate (MAC) operations performed by each layer,
as well as the amount of communication required to send and receive the
ciphertexts that contain each layer's inputs.
Figure~\ref{fig:layerstudy} shows the results of this study, plotting these microbenchmark convolution points, as well as each of the layers from VGG and from SqueezeNet. 
For the microbenchmark points, we varied image size from 2 to 32 by powers of
two, varied image channel values from 32 to 512 by powers of 2.  Following
the implementations of SqueezeNet and VGG16, we used filter sizes of 3 or 1. 
The data show that workloads like VGG (which are likely to see its same
energy benefits) are ones that maximize the number of MACs per MB of
communication.  
Workloads like SqueezeNet (which are likely to see its break-even or costs)
are ones that have fewer MACs per MB of communication.

These data provide two main benefits in interpreting \sys.  First, the data
show that a quick analytical comparison of computation (MACs) versus
communication (MBs) per layer helps an application designer decide if their
DNN application will see an energy benefit in the \sys client-aided model.
Second, the data point to an opportunity for future work, optimizing DNN
structure to maximize compute per communication for the \sys model.
\subsection{\sys with Model Privacy}
\sys targets a use case with relaxed model security requirements, and optimizes for minimal client-side computation and communication. However, when model security is a strict necessity, many optimizations presented would also benefit a client participating in a hybrid HE-MPC protocol. Namely, rotational redundancy can be applied to any HE algorithms using windowed rotations to reduce noise growth and enable smaller parameter selections, thus providing similar improvements in computation and communication to other protocols. Additionally, in any client-aided model encryption and decryption will repeatedly fall along the critical path. Therefore, hardware acceleration such as that presented in \arch will continue to be beneficial in reducing the time and energy required for these critical operations.

%% file: table/networks.tex
\begin{table}[!ht]
\renewcommand{\tabcolsep}{1pt}
\caption{Neural Networks used for system evaluation}
\label{tab:networks}
\centering
\begin{tabular}{c|c c c c|c|c c c|c c|c}
\hline
Network &\multicolumn{4}{c|}{\# Layers} & MACs & \multicolumn{3}{c|}{\% Acc. } & \multicolumn{2}{c|}{Mod. Sz. (MB)}& Comm.\\
 & Cnv & FC & Act & Pl & ($\times 10^{6}$) & Float & 8b & 4b & Float & 4b & (MB)  \\
\hline
LeNetSm\cite{mlpackkaggle} & 2 & 1 & 2 & 2  & 0.24 & 99.0 & 94.9 & 93.8 & 0.02& 0.01& 0.66 \\
\hline
LeNetLg \cite{TensorflowLenet}& 2 & 2 & 3 & 2 & 12.27& 98.7 & 97.2 & 96.4 & 8.22& 2.07& 2.6\\
\hline
SqzNet \cite{tensorsandbox} & 10 & 0 & 10 & 3  & 32.60 & 76.5 & 74.0 & 15.0 & 0.57& 0.16& 13.8\\
\hline
VGG16 \cite{VGGforCIFAR} & 13 & 2 & 14 & 5  & 313.26 & 70.0 & 66.0 & 21.0 & 56.40&  14.13& 22.2\\
\hline
\end{tabular}
\end{table}


%% file: src/relatedwork.tex
\section{Related Works}\label{sec:rw}

\subsection{Privacy Preserving DNN Inference}\label{sec:rw:ppdnn}
ML offloading requires data privacy. Recent work optimized server-centric
metrics, including usability \cite{CryptoNets, CHET}, training \cite{SecureML}, throughput (via batching)
\cite{CryptoNets, DNNTexas, nGraphHE2}, latency (via packing) \cite{LoLa,
CHET, Chameleon, Gazelle}, network complexity \cite{Gazelle, nGraphHE2,
MiniONN}, performance \cite{heax, HEAWS, heat}, and model privacy \cite{MiniONN, Gazelle, Cheetah, EzPC, Delphi, Soteria, XONN}. 
Unlike prior work that focused on the server, to the best of our knowledge,
\sys is the first work optimizing for resource-constrained client devices in client-aided HE. 

%
\subsection{HE Hardware Support}\label{sec:rw:fhehw}

Some prior work used hardware to accelerate kernels for lattice-based cryptography schemes~\cite{RLWEproc, RLWEsoftencrypt, VLSIRLWEproc, Ntt1}, including current state-of-art schemes~\cite{bfv1, bfv2, gentryFHE, ckks}.  Some directly accelerate HE~\cite{heax, VLSIencrypt, HEAWS, bootstrapNTT}, focusing primarily on hardware NTT. As we show in Figure \ref{fig:clientwntt}, NTT acceleration helps but is insufficient. Our work is the first to {\em comprehensively} optimize HE cryptographic primitives, which is crucial in client-aided HE. Furthermore, unlike prior work targeting large, high-power GPUs \cite{cuHE, FVgpu, RNSBFVgpu} and FPGAs \cite{heax, HEAWS, VLSIencrypt, heat}, \sys opts for an ASIC implementation, directly addressing the need for low-power, energy-efficient operation at the client device.

\subsection{Hardware Security}\label{sec:rw:hwsecurity}
Recent architectures offer privacy-preserving offloaded computation. 
Some techniques ensure data privacy, such as Trusted Execution Environments (TEEs)
\cite{SGX, nestedSGX, SecureEnclaves} and memory access
control and obfuscation \cite{ORAMdesign, byteblacklist, sDIMM, DATS}. While these prior techniques are vulnerable to side channel attacks, HE is not and data remain private while offloaded; HE is favorable thanks to its strong, proven privacy guarantee. Moreover, client-aided HE allows interactions between the client and server that are not allowed by TEEs~\cite{SGX}. 

\subsection{Low-Power ML Acceleration}\label{sec:rw:mlaccel}

Client DNN inference performance is improving through
software~\cite{Tensorflow2.2, PyTorch} and hardware
optimization~\cite{Eyeriss, EIE}.  One alternative to HE for private inference
is to simply outfitting IoT devices with local ML acceleration and computing
locally.  
However, as we argue in Section~\ref{sec:background}, local compute imposes
tight resource limits and requires maintaining (i.e., updating) models on a
potentially very large number of client devices, rather than an offload
server's centralized model. 
In contrast, \sys targets encrypted offload of ML (and other) computations, 
imposing few restrictions on centrally managed models.
Furthermore, \sys's support straightforwardly generalizes beyond ML: outfitting a device with a HE cryptographic acclerator, rather specialized DNN hardware, enables participating in 
\emph{any} client-aided, encrypted computation, not only ML. 
Encrypted applications research is an active, emerging area \cite{cryptorec,
SecureML, HEsort, HEsort2, encryptQueries}; \arch benefits a broad set of
existing and future encrypted applications. 

%% file: src/conclusion.tex
\section{Conclusion}\label{sec:conclusion}
In this work we present \sys, \syslong, a client-optimized system for privacy-preserving collaborative computing that enables participation from resource-constrained IoT client devices. We show that selecting efficient encryption parameters is critical to the performance of such a system and present rotational redundancy as an encrypted algorithm optimization to allow for more efficient selections. Because of its ability to use smaller ciphertexts, \sys reduces communication overheads by up to three orders of magnitude over existing privacy-preserving DNN inference protocols. Motivated by the remaining client computation bottleneck, we introduce \arch, hardware support to accelerate HE encryption and decryption along the critical path. By exploiting parallelism and supporting local data storage, \arch boasts a \encspeedup speedup and a \encenergysave energy savings for a single encryption operation. When integrated back into the full \sys system this translates to a \avgnnspeedup speedup on average for the client-side compute of DNN inference. For a reference implementation using Bluetooth communication, the combined hardware and software benefits from \arch make collaborative encrypted computing competitive against local compute with TFLite. A previously insurmountable task, this work, through its intentional client-aware optimizations, demonstrates that participation from resource-constrained IoT clients in collaborative encrypted computing is both feasible and even favorable.